\documentclass[a4paper,twoside,reqno]{bjp}
\usepackage{graphics}
\usepackage{cite}
\usepackage{amssymb,amsmath,amscd,amsthm}
\usepackage{times}

\usepackage[bookmarks=false]{hyperref}
\hypersetup{%
    colorlinks=true,        
    linkcolor=blue,          
    citecolor=blue,         
    urlcolor=blue           
    }

\usepackage{geometry}
\allowdisplaybreaks

\begin{document}

\title{Pre-outburst signal in the light curves of the recurrent novae RS~Oph and  T~CrB}

\runningheads{Zamanov, Marchev, Marti, Latev}{The recurrent novae RS~Oph and T~CrB }

\begin{start}{%
\author{R. K. Zamanov}{1},
\author{V. Marchev}{1},
\author{J. Marti}{2}, 
\author{G. Latev}{1}

\address{Institute of Astronomy and National Astronomical Observatory, Bulgarian Academy of Sciences, Tsarigradsko Shose 72, BG-1784 Sofia, Bulgaria}{1}
\address{Departamento de Fisica, Escuela Politecnica Superior de Jaen, Universidad de Jaen, Campus Las Lagunillas, A3, 23071, Jaen, Spain}{2}

\received{5 December 2023}
}

\begin{Abstract}
Pre-outburst signal (a decrease of the optical brightness)
just before the outburst is clearly detected in the observations of the T CrB 
obtained before and during the 1946 outburst. A similar decrease
is also visible in the light curve of RS Oph during the 2021 outburst.
We suppose that this is due to formation 
of a thick, dense envelope around the white dwarf, and we estimate 
its size (1000 - 2000 km), mass ($5 \times 10^{-8} - 6 \times 10^{-7}$ M$_\odot$) and 
average density (5 - 16 g cm$^{-3}$).   
\end{Abstract}

\begin{KEY}
stars (novae cataclysmic variables),  stars individual (RS Oph, TCrB)
\end{KEY}
\end{start}


\section{Introduction}
The Recurrent Novae (RNe) are classical novae that repeat their outbursts.
RNe are ordinary novae systems for which 
the recurrence time scale happens to be  
from a decade to a century. 
They are binary stars where matter accretes
from a donor star onto the surface of a white dwarf (WD), where
the accumulated material will start a thermonuclear explosion that makes the nova eruption (e.g. \cite{r01}, \cite{r02}). 
The two RNe discussed here (T~CrB and RS~Oph) belong to the group
of the RNe with red giant companions and with orbital periods of about one year, 
$P_{orb} = 227.6$~d for T~CrB \cite{r03}
and  $P_{orb} =  453.6$~d for RS~Oph \cite{r04}. 
T~CrB and RS~Oph are also classified as symbiotic stars, because the mass donor is 
a red giant. 
This type of nova is also referred to as a symbiotic recurrent nova  
(e.g. \cite{r05}, \cite{r06} ). 

In both stars a 
decrease of the B band brightness is observed a month before the nova outburst. 
In this work, we propose a hypothesis explaining
this drop of optical brightness.

\section{Pre-outburst signal - RS~Oph and  T~CrB}

\begin{figure}    
   \vspace{9.0cm}     
   \includegraphics{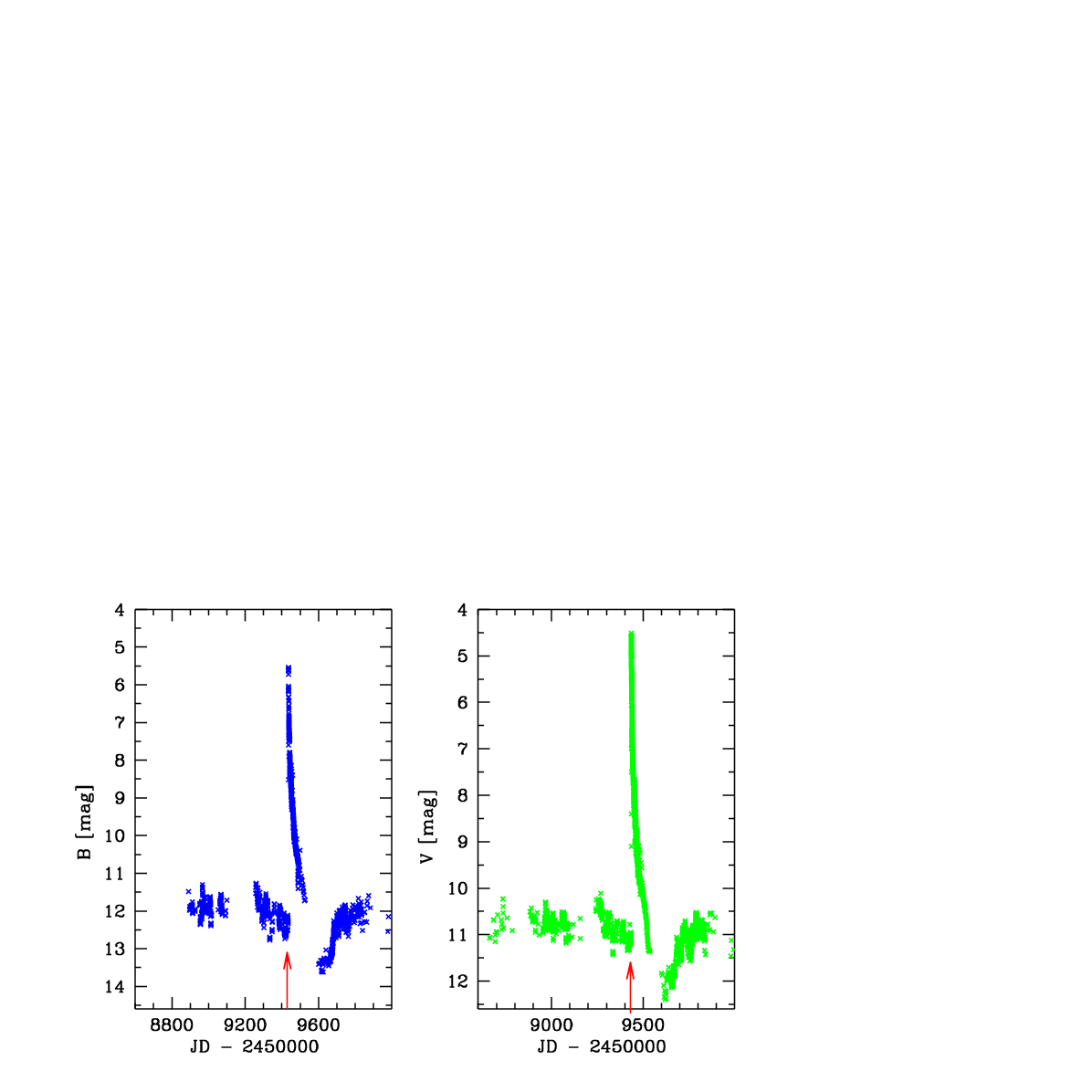}  
   \caption[]{AAVSO light curve of the recurrent nova RS~Oph around the 2021 outburst. 
                A drop of the brightness  
		before the 2021 outburst is visible in B as well as in V band. 
		The decrease of the brightness is with $\sim 0.5$ mag 
		and is indicated with red arrows.  
		}
   \label{f.RS} 
\end{figure}

Adamakis et al.\cite{r07} find a signal via wavelet analysis that 
can be used to predict a nova outburst. A drop in the B band magnitude  (decrease of the B band brightness with $\sim 1$~magnitude) 
just before the outburst is clearly detected in the photographic observations of the T CrB 
obtained before and during the 1946 outburst \cite{r08}. A similar decrease
(however with smaller amplitude) is visible in the light curve of RS Oph (Fig.~\ref{f.RS}).
Decrease of the mass accretion rate is the usual explanation 
for the brightness decrease of any accreting source 
and for T~CrB in particular (e.g. \cite{r09}, \cite{r10}). 
Here, we propose a different hypothesis, that the preoutburst decrease of the brightness
is due to formation of a dense envelope around the white dwarf. 

The accretion luminosity of an accreting white dwarf is:
\begin{equation}
L_{acc} = G \; \frac{M_{wd} \; \dot M_a}{R_{wd}},
\label{eq.r1}
\end{equation}
where G is the gravitational constant, $M_{wd}$ is the mass of the white dwarf,  $R_{wd}$ is its radius, 
$\dot M_a$ is the mass accretion rate. 
Our hypothesis is that a heavy (dense) envelope forms around the white dwarf.  
This dense envelope will later produce  TNR and  nova outburst. 
The envelope is impenetrable for the accreting matter and the $L_{acc}$  decreases:  
\begin{equation}
L_{acc} = G \; \frac{M_{wd} \; \dot M_a}{R_{wd} + \Delta R_{env}},
\label{eq.r2}
\end{equation}
where  $\Delta R_{env}$ is the size (thickness) of the envelope. 

\begin{figure}    
   \vspace{14.5cm}     
   \includegraphics{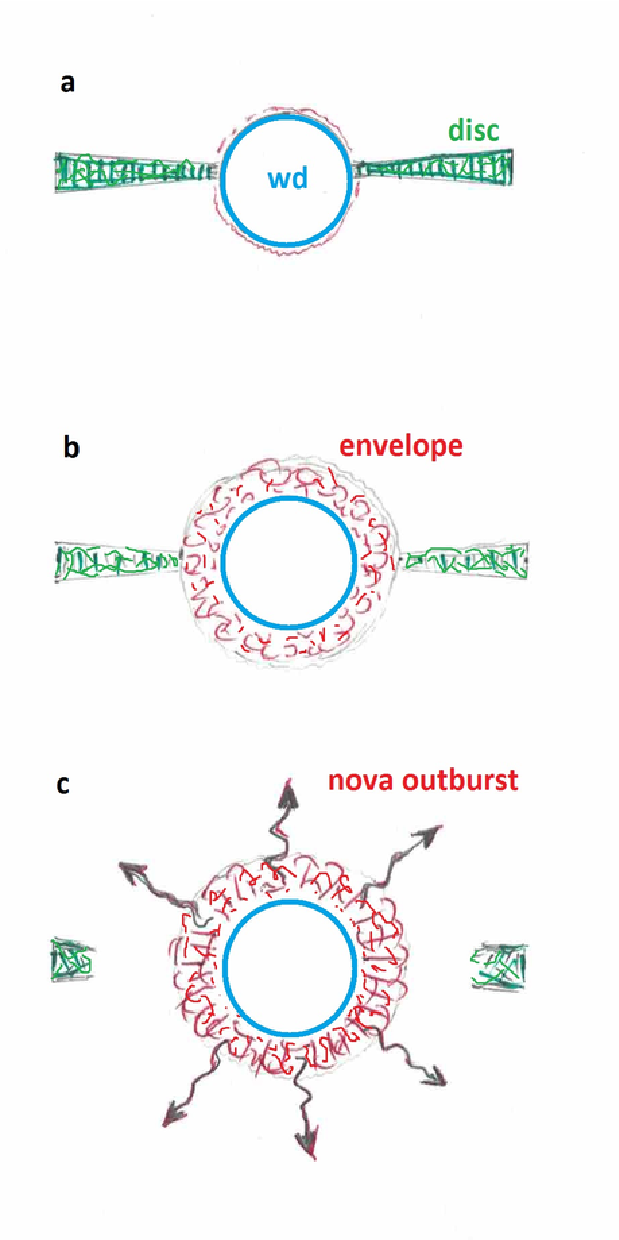}  
   \caption[]{A  sketch representing accreting white dwarf: 
    { \bf a)} the inner edge of the accretion disc reaches the surface of the white dwarf; 
    { \bf b)} a dense envelope forms and the inner edge of the accretion disc 
              is not able to go down to the surface of the white dwarf; 
    { \bf c)} the mass of the envelope exceeds the critical value and 
              produces a nova outburst.     
		}
   \label{f.sk} 
\end{figure}

A  sketch representing accreting white dwarf is drawn on Fig.~\ref{f.sk}.  
For most time of the outburst cycle  the envelope is thin and 
the inner edge of the accretion disc reaches the surface of the white dwarf (Fig.~\ref{f.sk}a). 
About 30-50 days before the outburst the envelope becomes thick and dense.
The inner edge of the accretion disc 
is not able to go down to the surface of the white dwarf. The brightness decreases (Eq.~\ref{eq.r2},  Fig.~\ref{f.sk}b). 
When the pressure exceeds the critical value, the white dwarf 
explodes as a nova (Fig.~\ref{f.sk}c). 

\vskip 0.3cm

{\bf RS~Oph:} The mass of the white dwarf in RS~Oph is estimated  $M_{wd} = 1.35 \pm 0.01$~M$_\odot$ 
on the basis of the  supersoft X-ray flux \cite{r11}. 
The  mass-radius relation for WDs gives  $R_{wd} =2296$~km, using the Eggleton's 
formula as given in \cite{r12}.
The ignition mass, $M_{ign}$, can be estimated from
\begin{equation}
P_{crit} = G \;  \frac{M_{wd} \; M_{ign}}{4 \pi \; R^4_{wd}}, 
\end{equation}
where $P_{crit}$ is the critical pressure for ignition, 
which is  $\approx10^{19}$ dyne cm$^{-2}$ (Jose et al. 2020).   
We estimate                     $M_{ign} = 9.7 \times 10^{-7}$~M$_\odot$,
   and average mass accretion rate   $\dot M_a = 4.8 \times 10^{-8}$~M$_\odot$~yr$^{-1}$,
   for a 20 years interval between the nova outbursts. 
Following Eq.~\ref{eq.r2}, for the B band brightness and $L_{acc}$ 
 to decrease by factor of 1.5, we estimate $\Delta R_{env} \approx 1150$~km.
 This corresponds to an average density in the envelope 16.2~g~cm$^{-3}$. 

\vskip 0.3cm

{\bf T~CrB:}  The mass of the white dwarf in T~CrB is estimated  $M_{wd} = 1.37$~M$_\odot$ 
on the basis of the radial velocities of the H$\alpha$ emission line \cite{r14}.
The  mass-radius relation for WDs gives  $R_{wd} =2018$~km. In the same way as above, 
we calculate $M_{ign} = 5.7 \times 10^{-7}$~M$_\odot$,
and average mass accretion rate   $\dot M_a = 7 \times 10^{-9}$~M$_\odot$~yr$^{-1}$,
for a 80 years interval between the nova outbursts. 
Following Eq.~\ref{eq.r2}, for the $L_{acc}$ and B band magnitude
to decrease by factor of 2, we estimate $\Delta R_{env} \approx 2000$~km.
This corresponds to an average density in the envelope 4.8~g~cm$^{-3}$, which is 
4 times denser than the water (and  slightly denser than the granite and aluminum). 

\vskip 0.3cm
Bruch \& Duschl\cite{r15} determined limits for the geometrical size of the boundary layer between 
the white dwarf and the accretion disc (T CrB and RS Oph are also included in their study)
and found typical values of $\gtrsim 2$ white dwarf radii. Our results indicate that 
the envelope is probably inside the boundary layer.  
It is worth noting, that Ilkiewicz et al. (2023) proposed that the super-active stage of 
T~CrB in the period 2015 -- 2023 is due to an activity
similar to disc instability of the dwarf novae. The disc 
instability can be the reason for the density enhancement of the  envelope. 
 

{\bf Conclusions: }
We suppose that the decrease of the optical brightness
before the nova outburst detected in the observations of the 
recurrent novae T CrB and RS Oph is a result of formation 
of a thick, dense envelope around the white dwarf. We estimate 
for this dense envelope size (1000 - 2000 km), 
mass ($5 \times 10^{-8} - 6 \times 10^{-7}$ M$_\odot$) and 
density (5 - 16 g cm$^{-3}$). 
The next outburst of T~CrB is expected soon and multifrequency observations 
can be valuable to understand the structure of the  envelope.

{\bf Acknowledgements: } 
We acknowledge project PID2022-136828NB-C42 funded by the Spanish 
MCIN/AEI/ 10.13039/501100011033 and "ERDF A way of making Europe". 
This research has made use of 
the AAVSO International Database contributed by observers worldwide.

\end{document}